\documentclass[a4paper,12pt]{elsarticle}
\usepackage[linesnumbered,boxed, noend, noline]{algorithm2e}
\usepackage{amsfonts}
\usepackage{txfonts} 
\pdfoutput=1

\newtheorem{thm}{Theorem}
\newtheorem{lem}[thm]{Lemma}
\newdefinition{rmk}{Remark}
\newproof{pf}{Proof}
\newproof{pot}{Proof of Theorem \ref{thm2}}
\newtheorem{propos}[thm]{Proposition}
\newtheorem{corollary}[thm]{Corollary}
\newtheorem{defn}[thm]{Definition}

\def\c{ , \hskip 3pt }
\def\lbrc{ \{ \smsp }
\def\rbrc{ \smsp \} }
\def\smsp{ \hskip 3pt }
\begin{document}
\begin{frontmatter}

\title{An n log n Algorithm for Deterministic Kripke Structure
Minimization}

\author{Karl Meinke, Muddassar A. Sindhu }

\address{School of Computer Science and Communication, Royal
Institute of
Technology 10044, Stockholm, Sweden.}

\ead{karlm@nada.kth.se, sindhu@csc.kth.se}
\begin{abstract}
We introduce an algorithm for the minimization of deterministic Kripke structures with 
$\mathcal O(kn  \log_{2} n)$ time complexity. We prove the correctness and complexity properties of this algorithm. 
\end{abstract}
\end{frontmatter}

\section{Introduction}
The problem of minimizing automata and transition systems has been widely studied in the literature. Minimization involves finding the smallest equivalent structure, using an appropriate definition of equivalence, (e.g. language equivalence or simulation equivalence).  In many software engineering applications, automata need to be minimized before complex operations such as model checking or test case generation can be carried out.

For different automata models and different notions of equivalence, 
 the complexity of the minimization problem can vary considerably.  The survey \cite{Ber11} considers minimization algorithms for DFA up to language equivalence, 
 with time complexities varying between $\mathcal O(n^{2})$ and $\mathcal O(n\;log\;n)$. 
Kripke structures represent a generalisation of DFA to allow non-determinism and multiple outputs. They
have been widely used to model concurrent and embedded systems. 
An algorithm for mimimizing Kripke structures has been given in
\cite{BusGru03}. 
In the presence of non-determinism, the complexity of minimization is quite high. 
Minimization up to language equivalence requires exponential time, while 
minimization up to a weaker simulation equivalence can be carried out in polynomial time
(see \cite{BusGru03}).

By contrast, we will show that {\em deterministic} Kripke structures can be efficiently minimized even up to language equivalence
with a worst case time complexity of $\mathcal O(kn \log_{2} n)$. For this, we generalise the concepts of 
right language and Nerode congruence from DFA to deterministic Kripke structures. We then show how the 
DFA minimization algorithm of  \cite{Hop71} can be generalised to compute the Nerode congruence $\equiv$
of a deterministic Kripke structure $\mathcal{K}$. 
The quotient Kripke structure $\mathcal{K} / \equiv$ is minimal and language equivalent to $\mathcal{K}$.
Our research \cite{MeinkeSindhu2011} into software testing has shown that this minimization algorithm 
makes the problems of model checking and test case generation more tractable for large models. 


The paper is organized as follows. In Section \ref{prem}, we introduce some mathematical pre-requisites. 
In Section \ref{sec3}, we give a minimization algorithm for deterministic Kripke structures. In Section \ref{sec4}, we give 
a correctness proof for this algorithm. In Section \ref{sec5} we provide a complexity analysis. 
Finally, in Section \ref{sec6} we discuss some conclusions.

\section{Preliminaries}
\label{prem}

We assume familiarity with the basic concepts of deterministic finite automata (DFA).
A \emph{Kripke structure} is a generalisation of a DFA to allow multiple outputs and non-determinism. 
A Kripke structure $\mathcal{K}$ over a finite set \emph{AP} of atomic propositions is a five tuple 
$\mathcal{K}=\langle Q,\Sigma,\delta,q_{0},\lambda\rangle$, where \emph{Q}, is the set of states, $\Sigma=\{\sigma_{1},...,\sigma_{n}\}$ is a finite alphabet, $\delta\subseteq Q\times\Sigma\times Q$ is the transition relation for states, $q_{0}$ is the initial state of $\mathcal{K}$ and $\lambda:Q\rightarrow2^{AP}$ is a function to label states. 
If $\vert \emph{AP} \vert = k$ we say that $\mathcal{K}$ is a $k$-bit Kripke structure.

We say that 
$\mathcal{K}$ is \emph{deterministic} if 
the relation $\delta$ is actually a function, $\delta: Q\times\Sigma\rightarrow Q$. We let $\delta^{*}:Q\times\Sigma^{*}\rightarrow Q$ denote the iterated state transition function where $\delta(q,\epsilon)=q$ and $\delta^{*}(q,\sigma_{1},...,\sigma_{n})=\delta(\delta^{*}(q,\sigma_{1},...,\sigma_{n-1}),\sigma_{n})$. Each property in \emph{AP }describes some local property of system states $q\in Q$. 
It is convenient to redefine the labelling function $\lambda$ as \emph{}$\lambda:Q\to\mathbb{B}^{k}$ given an enumeration of the set \emph{AP}. Then the iterated output function $\lambda^{*}:Q\;\times\;\Sigma^{*} \rightarrow \mathbb{B}^{k}$ is given by $\lambda^{*}(q,\sigma_{1},...,\sigma_{n})=\lambda(\delta^{*}(q,\sigma_{1},...,\sigma_{n}))$. More generally for any $q  \in Q$ define $\lambda_{q}^{*}(\sigma_{1},...,\sigma_{n})=\lambda^{*}(q,\sigma_{1},...,\sigma_{n})$. Given any $R\subseteq Q$ we write $\lambda(R)=\cup_{r\in R}\lambda(r)$. We let $q. \sigma$ denote $\delta(q,\sigma)$ and $R. \sigma$ denotes $\{r. \sigma \;\vert\; r \in R\}$ for $R\subseteq Q$.

\begin{figure}
\begin{centering}
\includegraphics[width=0.7\textheight,height=0.35\textheight]{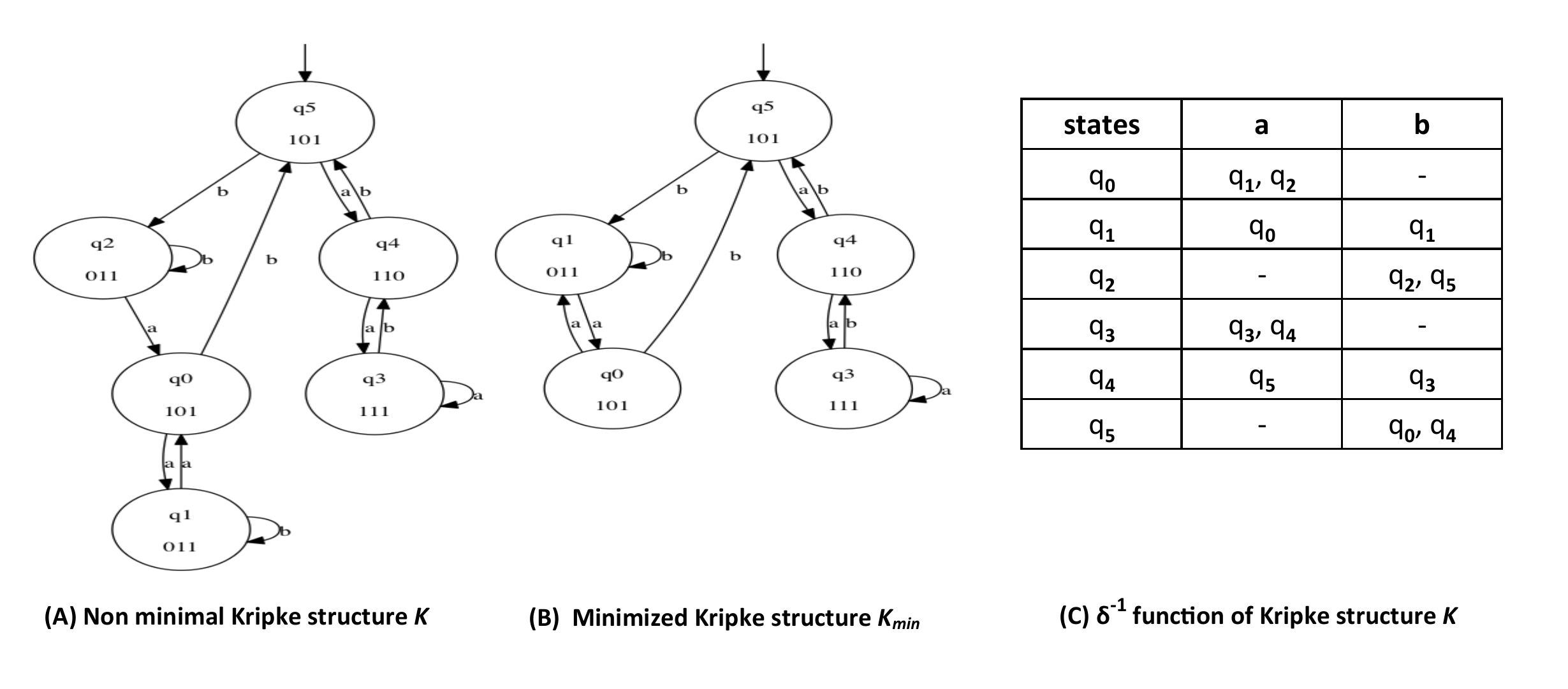}
\par\end{centering}

\caption{\label{fig:A-3-bit-Kripke} 3-bit Kripke Structure $\mathcal{K}$}
\end{figure}

We can represent a Kripke structure graphically in the usual way using a \emph{state transition diagram}. For example, a Kripke structure
with three bit labels in the output is shown in Fig \ref{fig:A-3-bit-Kripke}(A).

\subsection{Minimal DFA and minimal deterministic Kripke structures}

Let us consider a DFA $\mathcal{A}=\langle Q,\Sigma,\delta,q_{0},F\rangle$ . For each state $q\in Q$ of $\mathcal{A}$ there corresponds a
subautomaton of $\mathcal{A}$ rooted at $q$ which accepts the regular language 
$\mathcal{L}_{q}(\mathcal{A}) \subseteq \Sigma^*$, consisting of just those words accepted by the subautomaton with $q$ as initial state. 
Thus $\mathcal{L}_{q_0}(\mathcal{A})$ is the language accepted by $\mathcal{A}$.
The language $\mathcal{L}_{q}(\mathcal{A})$ is called either the \emph{future }of state \emph{q }or the \emph{right language }of \emph{q. }$\mathcal{A}$ is\emph{ minimal} if for each pair of distinct states \emph{$p,q\in Q$}, we have, $\mathcal{L}_{p}(\mathcal{A})\neq\mathcal{L}_{q}(\mathcal{A})$. For any regular language 
$\mathcal{L} \subseteq \Sigma^*$ there is a smallest DFA (in terms of the number of states) accepting $\mathcal{L}$. This DFA is minimal, and is 
unique up to isomorphism.

An equivalence relation $\equiv$ can be defined on 
the states of a DFA by $p\equiv q$ if and only if $\mathcal{L}_{p}(\mathcal{A})=\mathcal{L}_{q}(\mathcal{A})$.
This relation is a congruence, i.e. if $p \equiv q$  then 
$p. \sigma \equiv q. \sigma $ for all $\sigma \in\Sigma^{*}$. It is known as the \emph{Nerode congruence}. 
Consider the quotient DFA $\mathcal{A} / \equiv$. This is the unique smallest DFA 
which accepts the 
regular language $\mathcal{L}_{q_0}(\mathcal{A})$.
The problem of minimizing a DFA $\mathcal{A}$ is therefore to compute its Nerode congruence, which will be the identity relation if, and only if $\mathcal{A}$ is a minimal automaton.

The problem of computing a minimal Kripke structure $\mathcal{K}$  is 
an analogous but more general problem. 
In this case, the right language $\mathcal{L}_{q}(\mathcal{K})$ associated with a state $q$ of $\mathcal{K}$ can be defined by
$$
\mathcal{L}_{q}(\mathcal{K}) = 
\lbrc (\sigma_{1},...,\sigma_{n} , a ) \in \Sigma^* \times  \mathbb{B}^{k} 
\smsp \vert \smsp \lambda_{q}^{*}(\sigma_{1},...,\sigma_{n}) = a \rbrc .
$$
As before, $\mathcal{K}$ is\emph{ minimal} if for each pair of distinct states \emph{$p,q\in Q$} we have, $\mathcal{L}_{p}(\mathcal{K})\neq\mathcal{L}_{q}(\mathcal{K})$. 
There is again a smallest Kripke structure  associated with a right language 
$\mathcal{L} \subseteq \Sigma^* \times  \mathbb{B}^{k}$. This Kripke structure is also minimal, and 
unique up to isomorphism.
The Nerode congruence for a Kripke structure $\mathcal{K}$ is now defined by:
\begin{center}
$p\equiv q$ if and only if  $\lambda_{p}^{*}(\sigma_{1},...,\sigma_{n})=\lambda_{q}^{*}(\sigma_{1},...,\sigma_{n})$ for all $(\sigma_{1},...,\sigma_{n})\in\Sigma^{*}$.
\par\end{center}
and 
$\mathcal{K} / \equiv$ is the unique smallest Kripke structure 
associated with the 
right language $\mathcal{L}_{q_0}(\mathcal{K})$.
So the problem of minimising $\mathcal{K}$ is to compute this congruence.

\section{Kripke Structure Minimization
Algorithm}
\label{sec3}
	
    \begin{algorithm}\label{KMN}
    \caption{Kripke Structure Minimization}\label{alg1}
\KwIn{A deterministic Kripke structure $\mathcal{K}$ with no unreachable states and $k$ output bits.}
\KwOut{The Nerode congruence $\equiv$ for $\mathcal{K}$, i.e. equivalence classes of states for the 
minimized structure $ \mathcal{K}_{min}$ behaviourally equivalent to $\mathcal{K}$.}
     

Create\nllabel{partition} an initial state partition
$P=\{B_{q}=\{q^{\prime}\in
Q\;\vert\;\lambda(q)=\lambda(q^{\prime})\}\;\vert\;
q\in Q\}$. Let $n=\vert P\vert$. Let $B_{1},...,B_{n}$ be an
enumeration of P.

\textbf{if} {$n=\vert Q\vert$} \textbf{then}  \textbf{go to} line \ref{termination}.

    \ForEach{$\sigma\in\Sigma$\nllabel{subpartition}}{
\For{$i\leftarrow 1$ \KwTo $n$ } {$B(\sigma, i)=\{q\in B_{i}\;\vert\; \exists r \in Q \;s.t\; 
\delta(r,\sigma) = q\}.$
 /*This constitutes the subset of states in block $B_i$ which have predecessors
through input $\sigma$. */}
   
    }
$count=n+1$;

 \ForEach{$\sigma\in\Sigma$}{\nllabel{initwait}
choose\nllabel{initwait1} all the subsets $B(\sigma, i)$ (excluding any empty subsets) 
and put their block numbers $i$ on a waiting list (i.e. an
unordered
set) $W(\sigma)$ to be processed.
}
\BlankLine
Boolean splittable = true;

\While{splittable}{\nllabel{while}
 \ForEach{$\sigma\in\Sigma$ }{

 \ForEach{$i\in$ $W(\sigma)$}{

 Delete i from $W(\sigma)$\nllabel{bodystart}

\For{$j\leftarrow 1$ \KwTo $count-1$ s.t. $\exists t \in B_{j}$
with $\delta(t,\sigma)\in B(\sigma,i)$}{
Create $B_{j}^{\prime}=\{t\in B_{j}\;\vert\;\delta(t,\sigma)\in
B(\sigma,i)\}$\nllabel{newBj}

\If{$B_{j}^{\prime}\subset B_{j}$}{
$ B_{count}=B_{j}-B_{j}^{\prime}$; \nllabel{block}   $ B_{j}=B_{j}^{\prime}$

\ForEach{$\sigma\in\Sigma$}{
$B(\sigma,count)=\{q\in  B(\sigma,j) \;\vert\;    q \in B_{count}  \}$;\nllabel{newacount}

$ B(\sigma,j)=\{q\in B(\sigma,j)
  \;\vert 
\;  q \in B_{j}
\}$ \nllabel{newaj}

\If{$ j\notin W(\sigma)$ and\nllabel{if} $0<\vert B(\sigma,j)\vert\le\vert B(\sigma,count)\vert$}{$ W(\sigma)=W(\sigma)\cup\{j\}$}
\Else{ $W(\sigma)=W(\sigma)\cup\{count\}$\nllabel{endif}}

}

$ count=count+1$;
}

}
}
}

splittable = false;

\ForEach{$\sigma\in\Sigma$}{
\If {$W(\sigma) \not= \emptyset$}{splittable=true; \nllabel{bodyend}}
}
}
Return \nllabel{termination}partition blocks
$B_{1},...,B_{count}$.
    \end{algorithm}

Algorithm \ref{KMN} presents an efficient algorithm to compute the Nerode congruence $\equiv$ of a 
deterministic Kripke structure $\mathcal{K}$, which is the same as the state set of the associated quotient 
Kripke structure $\mathcal{K} / \equiv$.
We demonstrate the behavior of this algorithm on a simple example  
given in Fig.\ref{fig:A-3-bit-Kripke}(A) as follows. 

The algorithm begins by inverting the state
transition table as shown in Fig.\ref{fig:A-3-bit-Kripke}(C). Then it creates four
initial blocks of states on the basis of unique bit labels which are:
$B_{1}=\{q_{0},q_{5}\}$, $B_{2}=\{q_{1,}q_{2}\}$,
$B_{3}=\{q_{3}\}$
and $B_{4} = \{q_{4}\}$. Next it is checked whether the number of blocks is equal to the number of states $\vert Q\vert$ 
of the given Kripke structure. This is not the case,
so the next step is to refine each partition block $B_{i}$ into subsets $B(\sigma , i)$ of states which have predecessors 
via each input symbol of $\sigma \in \Sigma$.
This gives $B(a,1)=\{q_{0}\}$, $B(b,1)=\{q_{5}\}$, $B(a,2)=\{q_{1}\}$,
$B(b,2)=\{q_{1},q_{2}\}$, $B(a,3)=\{q_{3}\}$, $B(b,3)=\{q3\}$ ,
$B(a,4)=\{q_{4}\}$ and $B(b,4)=\{q_{4}\}$. The next step is to
initialize the waiting list $W(\sigma)$ for each symbol $\sigma \in \Sigma$ by
inserting the block numbers of all non-empty subpartition blocks $B(\sigma , i)$ created in the previous
step. We obtain $W(a)=\{1,2,3,4\}$ and $W(b)=\{1,2,4\}$. 

Now the algorithm can refine the initial partition
$B_{1} , \ldots , B_{4}$
by iterating the loop on line \ref{while} until $W(\sigma)=\emptyset$ for all
$\sigma\in\Sigma$.
For $i=1$ and $a\in\Sigma$ we have $W(a)=\{2,3,4\}$ and 
$B(a,1)=\{q_{0}\}$. We can see that $\delta(q_{1},a)=q_{0}\in
B(a,1)$
and $\delta(q_{2},a)=q_{0}\in B(a,1)$. But both $q_{1}$ and
$q_{2}$ are
in $B_{2}$. Therefore $B_{2}^{\prime}\not\subset B_{2}$ and
hence
no refinement of the partition is possible in this step. 

We proceed
with the next iteration of the loop by deleting $i=2$ from
$W(a)$ so that $W(a)=\{3,4\}$. Now we have
$B(a,2)=\{q_{1}\}$.
We can see that $\delta(q_{0},a)=q_{1}\in B(a,2)$. Therefore we have $B_{1}^{\prime}=\{q_{0}\}$. Since $B_{1}^{\prime}\subset B_{1}$ we therefore split $B_{1}$ into
$B_{5}=B_{1}-B_{1}^{\prime}=\{q_{0}, q_{5}\}-\{q_{0}\}=\{q_5\}$ and $B_{1}=B_{1}^{\prime}=\{q_{0}\}$.
Next we update the subsets $B(\sigma, i)$ and we get
$B(a,1)=\{q_{0}\}$,
$B(b,1)=\{\}$, $B(a,5)=\{\}$ and $B(b,5)=\{q_{5}\}$. The
updated
waiting sets are then $W(a)=\{1,3,4\}$ and $W(b)=\{1,2,4,5\}$. Next
we choose $i=1$, $\sigma=a$ and $W(a)=\{3,4\}$ and we obtain
$B(a,1)=\{q_{0}\}.$
It can be seen that $\delta(q_{1},a)=q_{0}\in a(a,1)$ and
$\delta(q_{2},a)=q_{0}\in a(a,1)$.
Therefore $B_{2}^{\prime}=\{q_{1},q_{2}\}$, but
$B_{2}^{\prime}\not\subset B_{2}$
and hence no refinement of the partition is possible in this case. 
We delete $i=3$ from $W(a)$ and obtain $W(a)=\{4\}$ and $B(a,3)=\{q_{3}\}$. 
We then find that for $q_{4}\in B_{4}$, $\delta(q_{4},a)=q_{3}\in B(a,3)$. Therefore we have $B_{4}^{\prime}=\{q_{4}\}$. But $B_{4}^{\prime}\not\subset B_{4}$, so no refinement of the partition is possible in this case. Continuing in the same way it will be seen that there is no further refinement of the partition possible for $i=4$ and $\sigma=a$ and for $i=1,2,4,5$ and $\sigma=b$ both $W(a)$ and $W(b)$ become empty. We terminate with five blocks in the partition. These constitute the states of our minimized Kripke structure as shown in Fig \ref{fig:A-3-bit-Kripke}(B).


%

\section{Correctness of Kripke Structure Minimization}
\label{sec4}
In this section we give a rigorous but simple proof of the correctness of Algorithm \ref{KMN}.
By means of a new induction argument, we have simplified
the correctness argument compared with \cite{Ber11} and \cite{Hop71}.
First let us establish termination of the algorithm by using an appropriate well-founded ordering for the main loop variant.

\begin{defn}
\label{def:definition3}Consider any pair of finite sets of
finite sets $A=\{A_{1},...,A_{m}\}$
and $B=\{B_{1},...,B_{n}\}$. We define an ordering relation
$\leq$ on $A$ and $B$
by $A\leq B$ iff $\forall1\leq i\leq m$, $\exists1\leq j\leq n$
such that $A_{i}\subseteq B_{j}$. Define $A<B\iff A\leq B\;\&\;
A\neq B$.
Clearly $\leq$ is a reflexive, transitive relation. Furthermore $\leq$ is well-founded, i.e. 
there are no infinite
descending chains $A_{1}>A_{2}>A_{3}...$ , since $\emptyset$ is
the smallest element under $\leq$.\end{defn}

\begin{propos}
Algorithm \ref{KMN} always terminates.
\end{propos}
\begin{pf}
We have two cases for the termination of the algorithm as a result of the partition formed on line \ref{partition} of the algorithm: (1) when $n=\vert Q\vert$, and (2) when $n<\vert Q\vert$.

 Consider the case when $n=\vert Q\vert$ then each block in the partition corresponds to a state of the
given Kripke structure with a unique bit-label and hence in this case the algorithm will terminate on line \ref{termination} by providing the description of these blocks.

Now consider the case when $n<\vert Q\vert$. Then the waiting sets $W(\sigma)$ for all $\sigma\in\Sigma$ will be initialized on lines \ref{initwait}, \ref{initwait1} and the termination of the algorithm depends on proving the termination of the loop on line \ref{while}. Now $W(\sigma)$  is intialized by loading the block numbers of the split sets on line \ref{initwait1}. There are only two possiblities after any execution of the loop. Let $W_{m}(\sigma)$ and $W_{m+1}(\sigma)$ represent the state of the variable $W(\sigma)$ before and after one execution of the loop respectively at any given time.  Then either $W_{m}(\sigma) = W_{m+1}(\sigma)\cup\{i\}$
and no splitting has taken place and \emph{i} is the deleted block number, or $W_{m}(\sigma)\cup\{j\}=W_{m+1}(\sigma)\cup\{i\}$ or $W_{m}(\sigma)\cup\{k\}=W_{m+1}(\sigma)\cup\{i\}$
where j and k represent the split blocks and one of them goes into $W_{m}(\sigma)$  if it has fewer incoming transitions. In either case $W_{m}(\sigma)>W_{m+1}(\sigma)$ by Definition \ref{def:definition3}. Therefore $W(\sigma)$ strictly decreases with each iteration of the loop
on line \ref{while}. Since the ordering $\leq$ is well-founded, Algorithm \ref{KMN} must terminate. \end{pf}

Now we only need to show that when Algorithm \ref{KMN} has terminated, it returns 
the Nerode congruence $\equiv$  on states. 

\begin{propos}
Let $P_{i}$ be the partition (block set) on the $ith$ iteration of Algorithm \ref{KMN}. For any blocks $B_{j}, B_{k} \in P_{i}$ and any states $p\in B_{j}, q\in B_{k}$  if $j\neq k$ then
$p\nequiv q$.\end{propos}

\begin{pf}
By induction on the number  $i$ of times the loop on line \ref{while} is executed. 
\vskip 6pt
\noindent\textbf{Basis:} Suppose $i=0$ then clearly the result holds because each block created at line \ref{partition} is distinguishable by the empty string $\epsilon$. 
\vskip 6pt
 \noindent\textbf{Induction Step:} Suppose $i=m>0$. Let us assume that the proposition holds after $m$ executions of the loop.

Consider any $B_{j},B_{k}\in P_{m}$.  During the $m+1$th execution of the loop on line \ref{while} either block $B_{j}$ is split into $B_{j}^{\prime}$ and $B_{j}^{\prime\prime}$  or  $B_{k}$ is split into $B_{k}^{\prime}$ and $B_{k}^{\prime\prime}$  but not both during one execution of the loop (due to line \ref{block}). 

Consider the case when $B_{j}$ is split then for any $p \in B_{j}$,  either $p \in B_{j}^{\prime}$ or $p \in B_{j}^{\prime\prime}$.  But for any $p \in B_{j}$ and $q \in B_{k}$, $p \nequiv q$ by the induction hypothesis. Therefore, for $p \in B_{j}^{\prime}$  or $p\in B_{j}^{\prime\prime}$  $p \nequiv q$. Hence the proposition is true for $m+1$th execution of the loop in this case. 

 By symmetry the same argument holds when $B_{k}$ is split.
\end{pf}

The following Lemma gives a simple, but very effective way to understand Algorithm \ref{KMN}.
Note that this analysis is more like a temporal logic argument than a loop invariant approach.
This approach reflects the non-determinism inherent in the algorithm.

\begin{lem}\label{lm}
For any states $p , q \in Q$, if $p \not\equiv q$ and initially $p$ and $q$ are in the same block 
$p , q \in B_{i_0}$ then eventually $p$ and $q$ are split into different blocks, 
$p \in B_j$ and $q \in B_k$ for $j \not= k$.

\end{lem}
\begin{pf}
Suppose that $p \not\equiv q$ and that initially $p , q \in B_{i_0}$ for some block $B_{i_0}$. 
Since $p \not\equiv q$ then for some $n \geq 0$, and 
$\sigma_1 , \ldots , \sigma_n \in \Sigma$, $$
\lambda^* ( p , \sigma_1 , \ldots , \sigma_n ) \not= 
\lambda^* ( q , \sigma_1 , \ldots , \sigma_n ) .
$$
We prove the result by induction on $n$.
\vskip 6pt
\noindent\textbf{Basis} Suppose $n = 0$, so that $\lambda ( p  ) \not= \lambda ( q  )$. By line  \ref{partition}, 
$p \in B_p$ and $q \in B_q$ and $B_p \not= B_q$. So the implication holds vacuously.
\vskip 6pt
\noindent\textbf{Induction Step} Suppose  $n > 0$ and for some $\sigma_1 , \ldots , \sigma_n \in \Sigma$,
$$
\lambda^* ( p , \sigma_1 , \ldots , \sigma_n ) \not= 
\lambda^* ( q , \sigma_1 , \ldots , \sigma_n ) .
$$
\noindent{\bf(a)} Suppose initially $\delta(p, \sigma_{1})\in B(\sigma_{1}, \alpha)$ and $\delta(q, \sigma_{1}) \in B(\sigma_{1}, \beta)$ for $\alpha \neq \beta$.

Consider when $\sigma = \sigma_{1}$ on the first iteration of the loop on line \ref{while}. Clearly, $B(\sigma_{1}, \alpha ), B(\sigma_{1}, \beta ) \in W(\sigma)$ at this point. Choosing $i= \alpha$ and $j=i_{0}$ on this iteration then since $\delta(p,\sigma_{1})\in B(\sigma_{1},\alpha)$ we have
\begin{center}
$B_{i_{0}}^{\prime}=\{t\in B_{i_{0}}\;\vert\;\delta(t,\sigma_{1})\in B(\sigma_{1},\alpha)\}\subset B_{i_{0}}$
\end{center}
This holds because $q \in B_{i_0}$ but  $\delta( q , \sigma_1 ) \in B( \sigma_1 , \beta )$ and 
$B( \sigma_1 , \alpha ) \not= B( \sigma_1 , \beta )$ so 
$B( \sigma_1 , \alpha ) \cap B( \sigma_1 , \beta ) = \emptyset$ and hence 
$q \not\in B'_{i_0}$.
Therefore $p$ and $q$ are split into different blocks on the first iteration so that 
$p \in B'_{i_0}$ and $q \in B_{i_0} - B'_{i_0}$.

By symmetry,  choosing $i = \beta$ and $j = {i_0}$ then $p$ and $q$ are split on the first loop iteration with 
$q \in B'_{i_0}$ and $p \in B_{i_0} - B'_{i_0}$.

\noindent{\bf (b)} Suppose initially 
$\delta( p , \sigma_1 ) , \delta( q , \sigma_1 ) \in B( \sigma_1 , \alpha )$
for some $\alpha$. Now 

\begin{center}
$
\lambda^* (\; \delta( p , \sigma_1 ) , \sigma_2 , \ldots , \sigma_n \; ) \not= 
\lambda^* ( \;\delta( q , \sigma_1 ) , \sigma_2 , \ldots , \sigma_n  \;) .
$
\end{center}

So by the induction hypothesis, eventually $\delta( p , \sigma_1 )$ and $\delta( q , \sigma_1 )$
are split into different blocks, $\delta( p , \sigma_1 )\in B_{\alpha}$ and 
$\delta( p , \sigma_1 )\in B_{\beta}$. At that time one of $B_{\alpha}$ or $B_{\beta}$
is placed in a waiting set $W(\sigma)$. Then either on the same iteration of the loop on line \ref{while} or on the 
next iteration, we can apply the argument of part (a) again to show that $p$ and $q$ are split into 
different blocks.
\end{pf}

Observe that only one split block is loaded into $W(\sigma)$
on lines \ref{if}-\ref{endif}.  From the proof 
of Lemma \ref{lm} we can see that it does not matter logically which of these two blocks we insert into $W(\sigma)$. However, 
by  choosing the subset with fewest incoming transitions we can obtain a worst case time complexity 
of order  $O( kn \smsp log_2 \smsp n )$, as we will show.

\begin{corollary}
For any states $p , q \in Q$, if $p \not\equiv q$ 
then $p$ and $q$ are in different blocks when the algorithm terminates.
\end{corollary}

\begin{pf}
Assume that $p \not\equiv q$. 

\noindent{\bf (a)} Suppose at line 3 that $n = \vert Q  \vert$. Then initially, all blocks $B_i$ are singleton sets and 
so trivially  $p$ and $q$ are in different blocks when the algorithm terminates.
\vskip 0.12pt

\noindent{\bf (b)} Suppose at line 3 that $n < \vert Q  \vert$. 

\noindent{\bf (b.i)} Suppose that  $p$ and $q$ are in different blocks initially. 
Since blocks are never merged then the result holds. 

\noindent{\bf (b.ii)} Suppose that  $p$ and $q$ are in the same block initially. 
Since $p \not\equiv q$ then the result follows by Lemma \ref{lm}.

\end{pf}
\section{Complexity Analysis}
\label{sec5}
Let us consider the worst-case time complexity of  Algorithm \ref{KMN}.
\begin{propos}
If $\mathcal{K}$ has $n$ states and $\Sigma$ has $k$ input symbols then Algorithm \ref{KMN}
has worst case time complexity $O( kn \log_{2} n )$.
\end{propos}
\begin{pf}
Creating the initial block partition on line \ref{partition} requires at most $O(n)$ assignments. 
The block subpartitioning in the loop on line \ref{subpartition} requires at most $O(kn)$ moves of states. Also the 
the initialisation of the waiting lists $W(\sigma)$ in the loop on line \ref{initwait}  requires at most $O(kn)$ assignments.

Consider one execution of the body of the loop starting on line \ref{while}, i.e. lines \ref{bodystart} - \ref{bodyend}. Consider any states 
$p \c q \in Q$ and suppose that $\delta ( p \c  \sigma ) = q$ for some $\sigma \in \Sigma$. Then the state $p$ 
can be: (i) moved into $B'_j$ (line \ref{newBj}), (ii) removed from $B_j$ (line \ref{block}), or (iii) moved into 
$B( \sigma \c  i)$ or $B( \sigma \c  count)$ (lines \ref{newacount}, \ref{newaj}) if, and only if, a block $i$ is being 
removed from $W(\sigma)$
such that $q \in B( \sigma \c  i)$ at that time. 
(Such a block sub-partition $B( \sigma \c  i)$ can be termed a {\it splitter} of $q$.)

Now each time a block $i$ containing $q$ is removed from $W(\sigma)$ its size is less than half of the size when it 
was originally entered into $W(\sigma)$, by lines \ref{if}-\ref{endif}. 
So $i$ can be removed from $W(\sigma)$ at most $O( log_2 \smsp n )$ times. 
Since there are at most $k$ values of $\sigma$ and $n$ values of $p$, then the total number of state moves 
between blocks and block sub-partitions is at most $O( kn \smsp log_2 \smsp n )$.
\end{pf}

\section{Conclusions}
\label{sec6}
We have given an algorithm for the minimization of deterministic Kripke structures with 
worst case time complexity $\mathcal O(kn \log_{2} n)$. We have analysed the correctness and performance 
of this algorithm. An efficient implementation of this algorithm has been developed which confirms the run-time performance theoretically predicted in Section 5.
This research has been supported by the Swedish Research Council (VR), the Higher Education Commission of Pakistan (HEC), as well as EU projects HATS FP7-231620, and MBAT ARTEMIS JU-269335.

\bibliographystyle{elsarticle-harv}
\bibliography{bibfile}

\end{document}